\documentclass[a4paper]{article}

\usepackage{INTERSPEECH2021}
\usepackage{subcaption}
\usepackage{multirow}
\usepackage{hyperref}

\title{Variational Information Bottleneck for Effective Low-resource Audio Classification}
\name{Shijing Si$^1$, Jianzong Wang$^{1\ast}$\thanks{$\ast$ Corresponding author: Jianzong Wang, jzwang@188.com}, Huiming Sun$^{1,2}$, Jianhan Wu$^3$, Chuanyao Zhang$^3$, Xiaoyang Qu$^1$, Ning Cheng$^1$, Lei Chen$^2$ and Jing Xiao$^1$}
\address{
  $^1$Ping An Technology (Shenzhen) Co., Ltd.\\
  $^2$Hong Kong University of Science and Technology\\
  $^3$University of Science and Technology of China}
\email{}

\begin{document}

\maketitle
\begin{abstract}
Large-scale  deep  neural  networks  (DNNs)  such  as  convolutional neural networks (CNNs) have achieved impressive performance in audio  classification for their powerful  capacity and strong generalization ability. However, when training a DNN model on low-resource tasks, it is usually prone to overfitting the small data and learning too much redundant information. To address this issue, we propose to use variational information bottleneck (VIB) to mitigate overfitting and suppress irrelevant information. In this work, we conduct experiments on a 4-layer CNN. However, the VIB framework is ready-to-use and could be easily utilized with many other state-of-the-art network architectures. Evaluation on a few audio datasets shows that our approach significantly outperforms baseline methods, yielding $\geq 5.0$\% improvement in terms of classification accuracy in some low-source settings.
\end{abstract}

\section{Introduction}

Deep learning \cite{lecun2015deep,schmidhuber2015deep,goodfellow2016deep} has emerged as the de facto standard technique in all areas of artificial intelligence, including speech processing \cite{hershey2017cnn,cho2018deep}, computer vision (CV) \cite{voulodimos2018deep,guo2020gluoncv} and natural language processing (NLP) \cite{young2018recent,otter2020survey}.
Currently DNNs have been very successful in the audio processing domain \cite{nam2018deep,purwins2019deep} for their strong capacity.
The DNN
models can produce good-quality representation features for audio
processing tasks, for example, audio classification \cite{massoudi2021urban}, automatic speech recognition \cite{moritz2020streaming} and speaker verification \cite{bai2020partial}. However, modern DNN-based audio classifiers typically require large amount of labeled data for training or fine-tuning \cite{wang2020few}, which might pose a challenge for many real applications. 
Because audio data labeling is both time-consuming and tedious, in many real situations there are only a limited number of training examples available \cite{georgiev2017low}. 

Applying DNN classifiers to low-resource datasets often leads to overfitting because DNNs have too much capacity and extract too many features of the low-resourced data that are irrelevant to the target labels \cite{pons2019training}. Variational information bottleneck (VIB) addresses the overfitting problem by eliminating irrelevant information and only retaining target related information \cite{alemi2017deep}.
Before elaborating VIB, we describe some basics on information bottleneck (IB).

IB was proposed by \cite{tishby2000information} to explain and enhance the generalization ability of neural networks. The main idea is: for an input data $X$ and its corresponding output (label) $Y$, we aim to learn a low-dimensional representation $Z$ that is maximally informative about our target $Y$ with minimal redundant information. IB method maximizes the mutual information $I(Z, Y)$ between $Y$ and $Z$, and minimizes the mutual information $I(Z, X)$ between $X$ and $Z$, so as to reserve the most useful data and discard redundant information.

The IB principle is appealing, since it defines what we mean by a good representation, in terms of the
fundamental tradeoff between having a concise representation and good predictive power \cite{tishby2015deep}. The main drawback of the IB principle is that mutual
information is, in general, computationally challenging. To address the computing issue, \cite{chalk2016relevant,alemi2017deep} proposed a variational inference approach, \emph{i.e.}, the VIB. It has been used in various areas of deep learning research. For example, \cite{mahabadi2021variational} proposed VIBERT that can suppress irrelevant features and yield a concise representation for text classification tasks in NLP.
\cite{dai2018compress} utilized the VIB principle to prune individual neurons for model compression. Though VIB has been shown great promise in low-resource text classification, few research have investigated its use in audio classification. 
In this paper, we propose to implement the VIB method to address the overfitting problem in low-resourced audio classification. To illustrate how the VIB framework works, we take convolutional neural networks (CNNs) as our feature extractor and conduct extensive experiments to verify its effectiveness. Actually, the VIB framework can also be used with other state-of-the-art network architectures, like Transformer encoder \cite{chi2021audio}, etc.

The main contributions are summarized as follows:
\begin{itemize}
    \item We explore the VIB approach to address the overfitting in audio classification
    \item We conduct extensive experiments to verify the effectiveness of VIB in comparison with other baselines
\end{itemize}

\section{Methodology}


The objective of IB is to learn a maximally compressed representation $Z$ of the input $X$ that maximally preserves information about the output $Y$. Its mathematical formula is shown in Eq. \eqref{eq:ib},
\begin{equation}\label{eq:ib}
    \mathcal{L}_{IB} = \underbrace{\beta I(X, Z)}_\text{Compression} -\underbrace{I(Z, Y)}_\text{Predictive} 
\end{equation}
 where $I(X, Z)$ and $I(Z, Y)$ are used for compression and predictive purposes, respectively, and
 $\beta\geq 0$ controls the balance between compression and prediction.

\begin{figure*}[!ht]
  \centering
  \includegraphics[width=\linewidth]{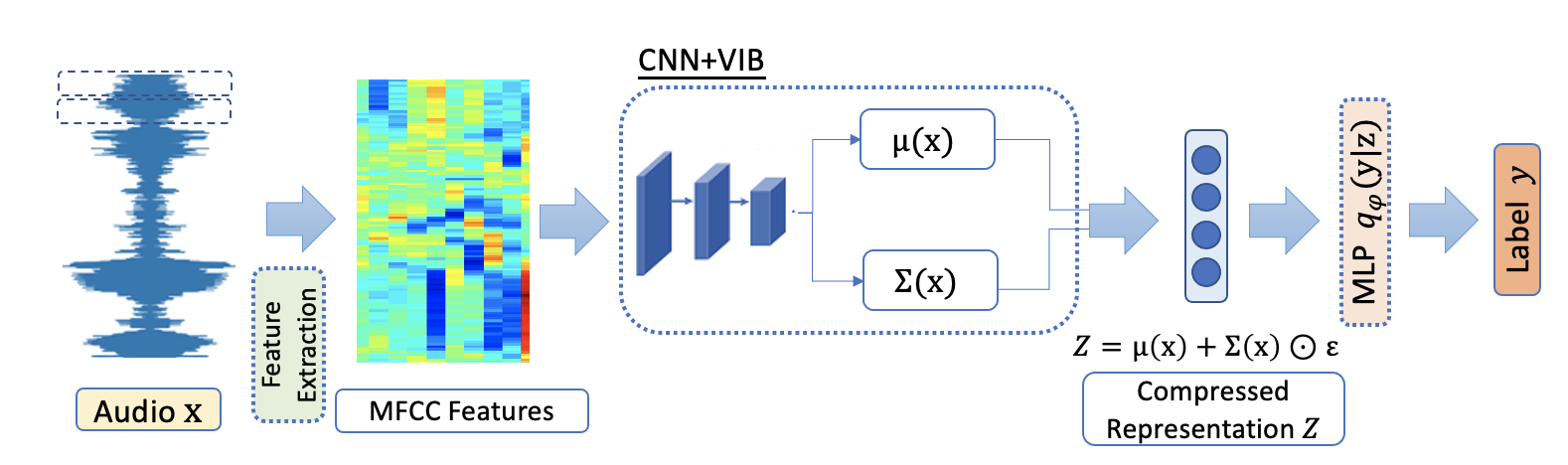}
  \caption{Schematic diagram of CNN+VIB framework.}
  \label{fig:cnn_vib}
\end{figure*}

\subsection{Variational Information Bottleneck}
Directly optimizing $\mathcal{L}_{IB}$ is hard, because it is usually computationally demanding. Then
\cite{chalk2016relevant,alemi2017deep} developed a variational approximate estimate of IB (VIB) in light of
\begin{equation}
\begin{split}
    & \beta I(X, Z) - I(Z, Y)\\
    &  \leq\beta\int dxdz{p}(x)p_{\theta}(z|x)\log\frac{p_{\theta}(z|x)}{r(z)}\\
    &~~-\int dxdydz{p}(x)p(y|x)p_{\theta}(z|x)\log{q}_{\phi}(y|z),
\end{split}
\end{equation}
where ${q}_{\phi}(y|z)$ is a variational distribution to approximate $p(y|z)$, and $r(z)$ is a specified prior distribution for latent code $Z$ (usually taking standard normal distribution), and $p_{\theta}(z|x)$ is an estimate of the posterior probability of $z$. 
Alternatively, $p_{\theta}(z|x)$ and $q_{\phi}(y|z)$ can also be interpreted as encoder and decoder, respectively as in the variational auto-encoder \cite{kingma13auto}.
For each training example $(x, y)$, the IB loss is upper bounded by
\begin{multline}\label{eq:vib_lb}
    \mathcal{L}_{VIB} = \beta \mathbf{E}_{x}[KL(p_{\theta}(z|x), r(z))] +\\ \mathbf{E}_{z\sim{p}_{\theta}(z|x)}[-\log{q}_{\phi}(y|z)].
\end{multline}
Therefore, in this work Eq. \eqref{eq:vib_lb}
is applied to minimize the IB objective.

\subsection{Deep VIB Audio Classifier}
Here we construct an audio classifier by incorporating a CNN encoder inside the VIB. The schematic diagram of our framework is shown in Fig. \ref{fig:cnn_vib}. Specifically, during training, for each audio data $x$, we extract its Mel-Frequency Cepstral Coefficients (MFCC), and pass them through a CNN encoder, yielding the mean and standard deviation vectors, \emph{i.e.}, $\mu(x)$ and $\Sigma(x)$, of the posterior distribution  $p_{\theta}(z|x)$. Then we simulate Gaussian samples of $z$ and feed them a multi-layered perceptron (MLP) $q_{\phi}(y|z)$ to predict the logits of label $y$. During inference, $z=\mu(x)$ is used instead of sampling from $p_{\theta}(z|x)$.

For each training example, $z$ is the latent bottleneck and its dimension $K$ controls the information relevant to the label $y$. If the dimension $K$ is large, the compression effect is decreased; otherwise the bottleneck compressed the information used for prediction.
The hyper-parameter $\beta$ in Eq. \eqref{eq:vib_lb} also controls the compression effect of the model. If $\beta$ is small, the compression effect is weak and the model tends to overfit the data. If $\beta$ is large, the model is compressed heavily and little information is used for prediction. In Fig. \ref{fig:cnn_vib}, CNN architecture could be replaced by other networks, like Transformer \cite{chi2021audio}, etc. In this work, we conduct experiments on a CNN architecture, because it is commonly used in audio classification.

\section{Experiment Setup}

Here we conduct audio classification experiments over multiple datasets to empirically verify the effectiveness of our method, in comparison with existing approaches.
We use Tensorflow and Keras to implement our VIB+CNN classifier
and Librosa \cite{mcfee2015librosa}
for audio processing and MFCC feature extraction.
We utilize the loss, accuracy and F1 score to evaluate the performance of different approaches.
We then present some ablation studies
and analyses to investigate the source of the improvements using our VIB method. The Python code for experiments can be found at \url{https://github.com/shijing001/VIB_audio_classification}.

\subsection{Datasets}
We conduct extensive experiments on
four datasets: Audio-MNIST, ESC-50, Toronto Emotional Speech Set (TESS), and TUT Acoustic Scenes (TUT). Details on these 4 datasets are presented as follows. Basic statistics, \emph{i.e.}, data size, number of classes ($C$), train/validation/test size and the number of samples per class, are shown in Table \ref{tab:data_info}. 


\noindent\textbf{Audio MNIST}
consists of 3000 audio recordings
of spoken digits (0-9) in English with 50 repetitions per digit for each of the 60 different speakers. Recordings were at a sampling frequency of 48kHz and were
saved in 16 bit integer format. 

\noindent\textbf{ESC-50} is a collection of short environmental recordings available in a unified format (5-second-long clips, 44.1 kHz, single channel, Ogg Vorbis compressed @ 192 kbit/s). It consists of a labeled set of 2000 environmental recordings (50 classes, 40 clips per class).

\noindent\textbf{Toronto Emotional Speech Set (TESS)}\cite{TESS2020} A set of 200 target words were spoken in the carrier phrase ``Say the word xxxx" by two actresses and recordings were made of the set portraying each of seven emotions (anger, disgust, fear, happiness, pleasant surprise, sadness, and neutral). There are 2800 recordings in total (7 classes, 400 clips per class).

\noindent\textbf{TUT} Acoustic Scenes 2017 consists of 4680 recordings of 15 classes. Each audio is about 10 seconds long. More details on this data can be found in \cite{DCASE2017challenge}.

\begin{table}[]
\caption{\label{tab:data_info}Basic statistics of four datasets with $C$ and Num/$C$ representing the number of labels and average number of examples for each class.}
\begin{tabular}{lllll}
\hline
Dataset     & Size & $C$ & Train/Valid./Test & Num/$C$ \\ \hline
Audio-MNIST & 3000 & 10      & 1800/600/600      & 300              \\
ESC-50      & 2000 & 50      & 1600/200/200      & 40               \\
TESS        & 2800 & 7       & 1680/560/560      & 400              \\
TUT         & 4680 & 15      & 3808/936/936    & 312              \\ \hline
\end{tabular}
\end{table}

\subsection{Deep Learning models}
We employ a 4-layered CNN architecture as the backbone in our experiments. It
consists of 4 convolutional layers with 32, 96, 96 and 160 output channels, respectively, followed by a maximum pooling layer and a fully connected (dense) layer with $C$ logits as output. Its architecture
is displayed in Table \ref{tab:cnn.arch}.
The baseline methods to address overfitting are listed as follows.

\begin{table}[]
\caption{The architecture of the baseline CNN classifier. \label{tab:cnn.arch}}
\centering
\begin{tabular}{llll}
\hline
Layer       & Outputs & Kernel & Stride \\ \hline
Conv2D+ReLU & 32      & $4\times4$    & 1      \\
Conv2D+ReLU & 96      & $4\times10$   & 1      \\
Conv2D+ReLU & 96      & $4\times10$   & 1      \\
Conv2D+ReLU & 160     & $4\times10$   & 1      \\
MaxPooling  & 160     & $2\times2$    & -      \\
Dense       & $C$       & -      & -      \\ \hline
\end{tabular}
\end{table}


\noindent\textbf{Weight Decay} is a common regularization technique to improve generalization \cite{krogh1991a}. It regularizes the large weights $w$ by adding a penalization term $\frac{\lambda}{2}\|w\|$ to the loss, where $\lambda$ is a hyperparameter specifying the strength of regularization. $\lambda$ is set to $1.0e-4$ in the experiments. 

\noindent\textbf{Dropout} \cite{srivastava2014dropout}, is a widely used stochastic regularization techniques used in deep learning models \cite{devlin2019bert,vaswani2017attention} to mitigate overfitting. We implement the spatial dropout for all 4 CNN layers with dropout probability $0.2$.

\noindent\textbf{CNN+VIB (Ours)} classifier is built on top of the baseline CNN model, and its architecture is the same as Table \ref{tab:cnn.arch} but with an additional dense layer after maxpooling, which yields the $\mu(x)$ and $\Sigma(x)$ for $p_{\theta}(z|x)$.
For the hyper-parameters $K$ and $\beta$, taking values in the ranges  $[20, 50, 100, 200]$ and $[2.0e-3, 5.0e-3, 2.0e-2, 5.0e-2]$, respectively. We first perform
model selection on the validation set to find the hyper-parameters and then evaluate the selected models
on the test set.

\subsection{Implementation Details}
\noindent\textbf{Preprocessing} All the raw audios are resampled to 44.1kHz and then fixed to the certain length by zero-padding or truncating (\emph{i.e.} 4s for the Audio-MNIST,
5s for the ESC-50, 2s for the TESS and 10s for the TUT). The short time Fourier transform (STFT) is then applied on the audio signals to calculate spectrograms, with a window size of 40ms and a hop size of 20ms. 40 mel filter banks are applied on the spectrograms followed by a logarithmic operation to extract the MFCC features.

\noindent\textbf{Training details} In the training phase, the Adam algorithm \cite{Kingma2015adam}
is employed as the optimizer with the default parameters. The
model is trained end-to-end with the initial learning rate of 0.001
and the exponential decay rate of 0.98 for each epoch. Batch size is set to 8 and training epoch is set to 20 for TUT and 40 for other datasets.

\section{Results and Analysis}

\begin{figure*}[h]
\begin{subfigure}{0.5\textwidth}
\includegraphics[width=0.99\linewidth]{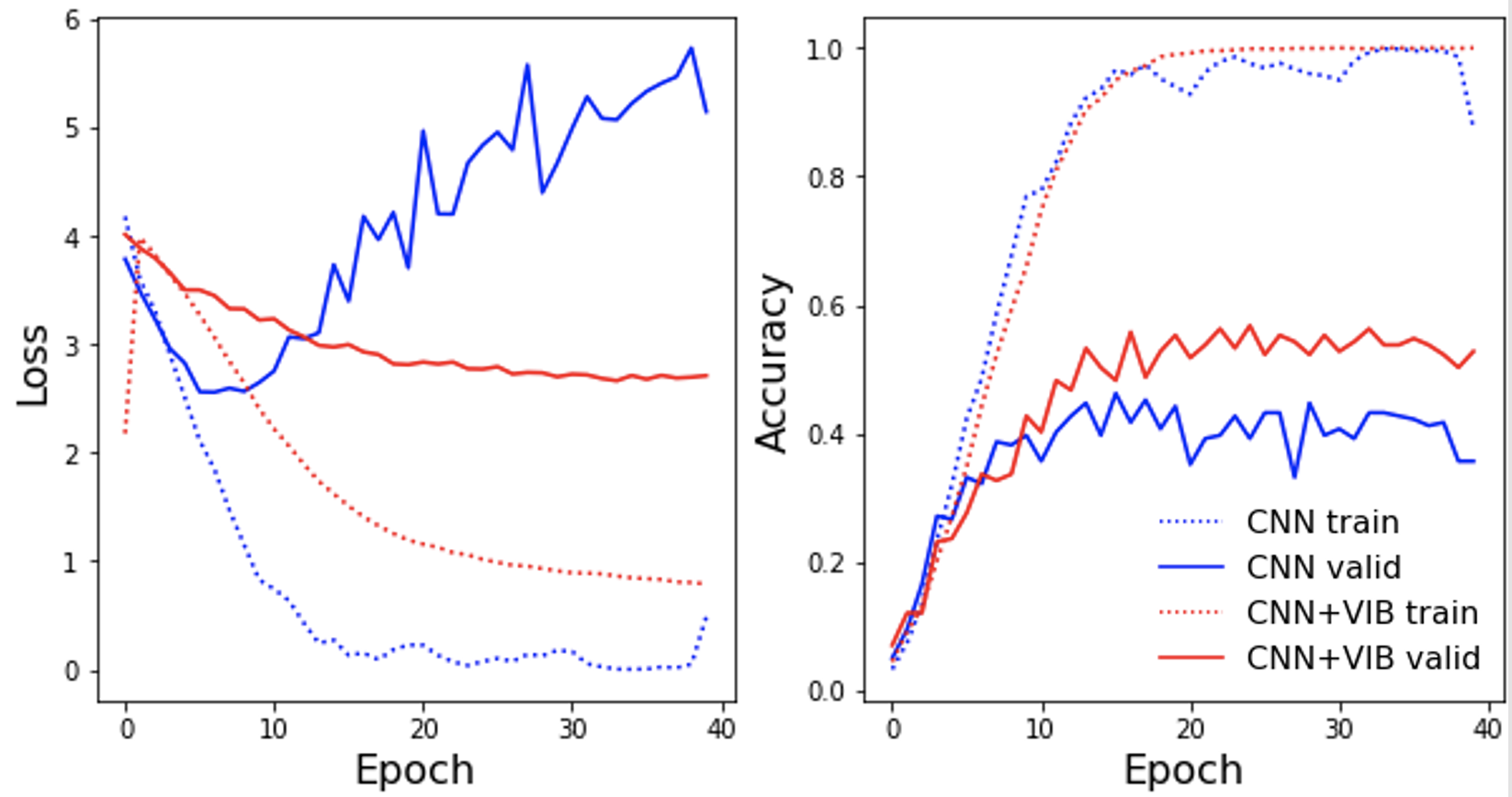} 
\caption{Training process of CNN and CNN+VIB on ESC-50 dataset}
\label{fig:subim1}
\end{subfigure}
\begin{subfigure}{0.5\textwidth}
\includegraphics[width=0.99\linewidth]{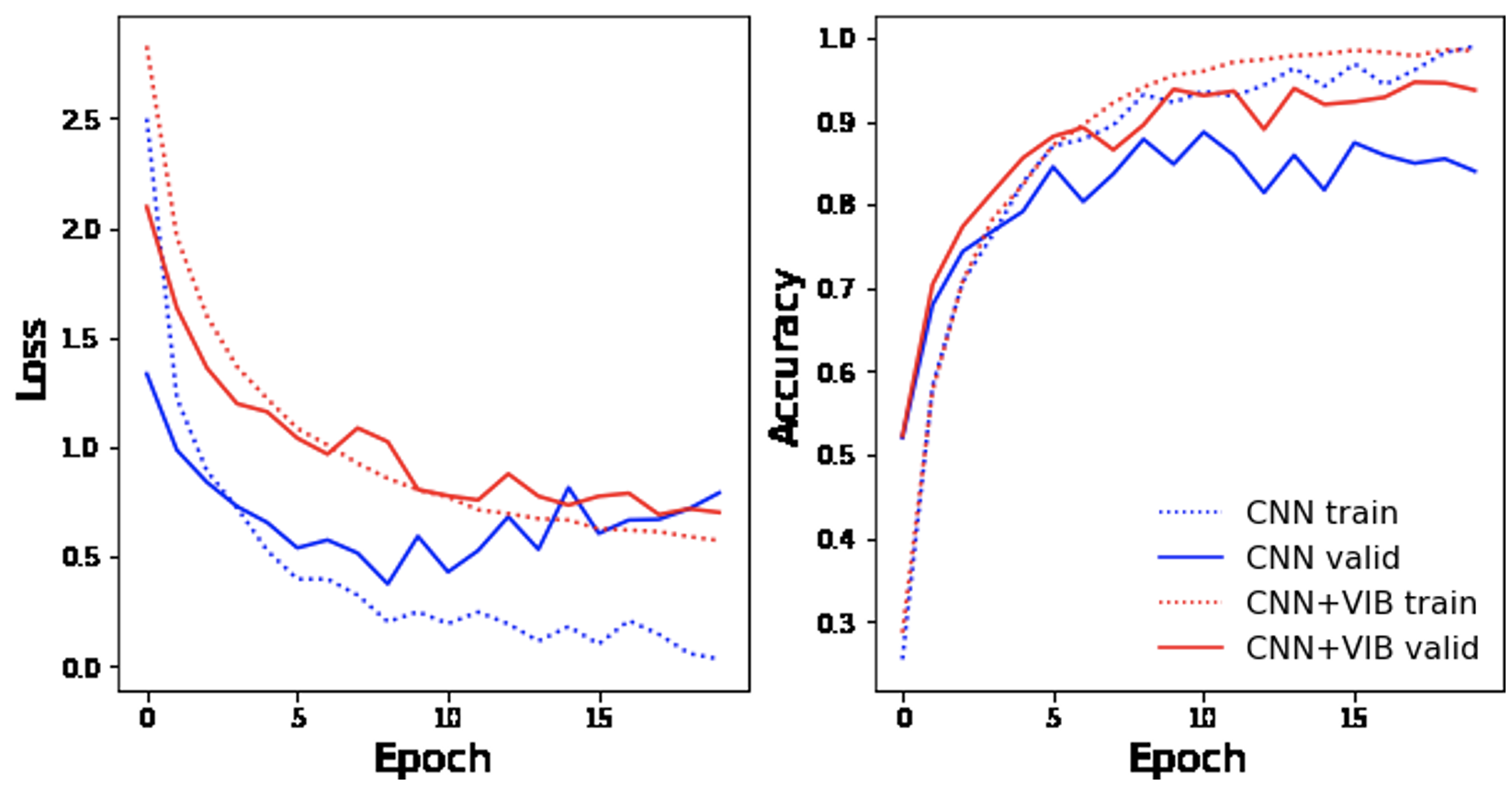}
\caption{Training process of CNN and CNN+VIB on TUT data}
\label{fig:subim2}
\end{subfigure}
\caption{The loss and accuracy of CNN+VIB (Red lines) versus CNN (Blue lines) on two datasets: ESC-50 and TUT. The dashed lines present the loss and accuracy on training sets, whereas the solid lines exhibit loss and accuracy on validation sets.}
\label{fig:training}
\end{figure*}

\subsection{Overfitting Suppression}

Here we exhibit that VIB can suppress overfitting during model training.
Fig. \ref{fig:training} summarizes the loss and accuracy on training and validation sets versus the number of training epochs. Fig. \ref{fig:subim1} and \ref{fig:subim2} present the performance on the ESC-50 and TUT datasets, respectively.
In Fig. \ref{fig:subim1}, the left plot illustrates the training and validation losses of CNN and CNN+VIB versus training epochs.
The blue dotted and solid lines represent the training and validation losses of the baseline CNN model, while the red dotted and solid lines indicate training and validation losses for our CNN+VIB approach. As the training epoch increases, the baseline CNN training loss (the blue dotted line) decreases quickly and then stabilizes after around 15 epochs. However, the baseline validation loss (the blue solid line) falls rapidly in the beginning, reaches its minimum at around 10 epochs and then increases speedily for the rest of training epochs. This is a clear sign that the baseline CNN model has overfitted the ESC-50 dataset. By contrast, the validation loss from CNN+VIB (red solid line) exhibit a gradually decreasing trend as the increase of training epochs. This is the empirical evidence of VIB reducing overfitting, which also appears in other datasets like the TUT in Fig. \ref{fig:subim2}.

The central-left plot of Fig. \ref{fig:training} displays the classification accuracy on training and validation sets of ESC-50. 
The blue dotted and solid lines represent the training and validation accuracy of the baseline CNN model, while the red dotted and solid lines indicate training and validation accuracy for our CNN+VIB approach. As the training epoch increases, the baseline CNN training accuracy (the blue dotted line) increases quickly and then stabilizes after around 15 epochs. However, the baseline validation accuracy (the blue solid line) grows rapidly in the beginning, reaches its maximum at around 15 epochs and then decreases slightly later.
Fig. \ref{fig:subim2} shows the loss and accuracy of CNN and CNN+VIB on the TUT dataset, which also exhibits the similar pattern.

\begin{table*}[]
\caption{\label{tab:performance}The performance (accuracy and F1 score) of four methods on audio-MNIST, ESC-50, TESS and TUT datasets under multiple low-resource settings that are characterized by varying the percentage of training data. Bold values indicate the best ones. The hyper-parameters are tuned by grid search.}
\begin{tabular}{l|l|lllllllllllllll}
\cline{1-16}
Data                                                                   & Model           & \multicolumn{2}{c}{5\%}         &  & \multicolumn{2}{c}{10\%}        &  & \multicolumn{2}{c}{30\%}        &  & \multicolumn{2}{c}{50\%}        &  & \multicolumn{2}{c}{100\%}       &  \\ \cline{1-4} \cline{6-7} \cline{9-10} \cline{12-13} \cline{15-16}
                                                                       &                 & Acc.           & F1             &  & Acc.           & F1             &  & Acc.           & F1             &  & Acc.           & F1             &  & Acc.           & F1             &  \\ \cline{1-16}
\multirow{4}{*}{\begin{tabular}[c]{@{}l@{}}audio-\\ MNIST\end{tabular}} & CNN             & 0.543          & 0.541          &  & 0.842          & 0.835          &  & 0.953          & 0.951          &  & 0.991          & 0.990          &  & 0.991          & 0.993          &  \\
                                                                       & +Dropout        & 0.572          & 0.572          &  & 0.824          & 0.821          &  & 0.952          & 0.951          &  & 0.991          & 0.991          &  & 0.990          & 0.990          &  \\
                                                                       & +Weight Decay   & 0.578 & 0.573 &  & 0.844          & 0.833          &  & 0.954          & 0.949          &  & 0.990          & 0.992          &  & 0.993          & 0.991          &  \\                               & \textbf{CNN+VIB} & \textbf{0.587} & \textbf{0.584}          &  & \textbf{0.868} & \textbf{0.869} &  & \textbf{0.975} & \textbf{0.975} &  & \textbf{0.997} & \textbf{0.997} &  & \textbf{1.000} & \textbf{1.000} &  \\ \cline{1-16}
\multicolumn{1}{c|}{\multirow{4}{*}{ESC-50}}                           & CNN             & 0.159          & 0.155          &  & 0.192          & 0.190          &  & 0.303          & 0.303          &  & 0.396          & 0.392          &  & 0.552          & 0.507          &  \\
\multicolumn{1}{c|}{}                                                  & +Dropout        & 0.162          & 0.161          &  & 0.213          & 0.207          &  & 0.295          & 0.287          &  & 0.502          & 0.436          &  & 0.551           & 0.503          &  \\
\multicolumn{1}{c|}{}                                                  & +Weight Decay   & 0.178          & 0.176          &  & 0.210          & 0.198          &  & 0.358          & \textbf{0.357} &  & 0.421          & 0.420          &  & 0.568          & 0.563          &  \\
\multicolumn{1}{c|}{}                                                  & \textbf{CNN+VIB} & \textbf{0.224} & \textbf{0.184} &  & \textbf{0.239} & \textbf{0.210} &  & \textbf{0.363} & 0.336          &  & \textbf{0.532} & \textbf{0.528} &  & \textbf{0.581} & \textbf{0.589} &  \\ \cline{1-16}
\multirow{4}{*}{TESS}                                                  & CNN             & 0.936          & 0.934          &  & 0.952          & 0.952          &  & 0.921          & 0.916          &  & 0.998          & 0.998          &  & 0.997          & 0.997          &  \\
         & +Dropout        & 0.938          & 0.937          &  & 0.967          & 0.964          &  & 0.991          & 0.991          &  & 0.993          & 0.993          &  & 0.997          & 0.997          &  \\
        & +Weight Decay   & 0.927          & 0.920          &  & 0.946          & 0.947          &  & 0.925          & 0.920          &  & 0.993          & 0.992          &  & 0.998          & 0.998          &  \\
 & \textbf{CNN+VIB} & \textbf{0.954} & \textbf{0.957} &  & \textbf{0.971} & \textbf{0.968} &  & \textbf{0.997} & \textbf{0.997} &  & \textbf{0.998} & \textbf{0.998} &  & \textbf{0.998} & \textbf{0.998} &  \\ \cline{1-16}
\multirow{4}{*}{TUT}                                                   & CNN             & 0.411          & 0.373          &  & 0.652          & 0.650          &  & 0.811          & 0.803          &  & 0.867          & 0.862          &  & 0.891          & 0.892          &  \\
                 & +Dropout        & 0.551          & 0.542          &  & 0.611          & 0.601          &  & 0.815          & 0.810          &  & 0.903          & 0.902          &  & 0.893          & 0.892          &  \\
                & +Weight Decay   & 0.571          & 0.553          &  & 0.623          & 0.617          &  & 0.834          & 0.831          &  & 0.869          & 0.864          &  & 0.927          & 0.925          &  \\
                                                                       & \textbf{CNN+VIB} & \textbf{0.607} & \textbf{0.595} &  & \textbf{0.674} & \textbf{0.670} &  & \textbf{0.855} & \textbf{0.857} &  & \textbf{0.912} & \textbf{0.913} &  & \textbf{0.942} & \textbf{0.943} &  \\ \cline{1-16}
\end{tabular}
\end{table*}

Table \ref{tab:performance} displays the performance (in terms of accuracy and F1 score) of four methods on audio-MNIST, ESC-50, TESS and TUT datasets under multiple low-resource settings. We create low-resource settings by subsampling the original training data with a certain percentage. The main findings of this table are i.) CNN+VIB can always outperform baseline methods in all settings. ii.) in extremely low-resource settings like only using 5\% training data, CNN+VIB outperforms the baseline methods by a significant margin and the margin decreases as the training data size increases.


\subsection{Ablation Study}

\begin{figure}[ht]
\begin{subfigure}{0.23\textwidth}
\includegraphics[width=0.99\linewidth]{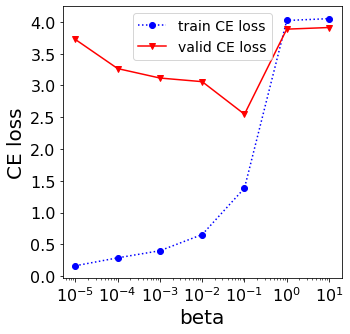} 
\caption{Fixed $K=20$ on ESC-50}
\label{fig:esc_beta}
\end{subfigure}
\begin{subfigure}{0.23\textwidth}
\includegraphics[width=0.99\linewidth]{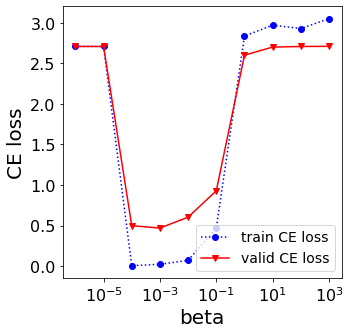}
\caption{Fixed $K=20$ on TUT data}
\label{fig:tut_beta}
\end{subfigure}
\caption{Validation and training cross-entropy losses of CNN+VIB for varying $\beta$ with $K$ fixed at 20.\label{fig:loss_beta}}
\end{figure}


To analyze the effect of VIB on reducing overfitting, we investigate the effect of the $\beta$ on training and validation cross-entropy (CE) losses since $\beta$ controls the trade-off between removing
information from the audio features (high $\beta$) and keeping information that is predictive of the target $y$
(low $\beta$). In Fig. \ref{fig:loss_beta}, we fix the bottleneck size ($K=20$) and train CNN+VIB
on two datasets (ESC-50 and TUT) for varying values of $\beta$ and plot the validation (red solid) and training (blue dotted) CE losses.
Fig. \ref{fig:esc_beta} displays a typical pattern of how training and validation losses evolve with the increase of $\beta$. For small values of $\beta$ (left side of Fig. \ref{fig:esc_beta}), where VIB has little effect, the validation loss is substantially higher than the training
loss, which indicates overfitting. This is because the network learns to be more deterministic ($\Sigma\approx 0$), thereby
retaining too much irrelevant information. As we increase $\beta$, where VIB has an effect, we observe better generalization performance with less overfitting. As $\beta$ becomes too large (right side of the plot), both the training and validation CE losses
shoot up because the amount of preserved information is insufficient to differentiate between the classes.
Fig. \ref{fig:tut_beta} shows a slightly different case. When $\beta$ is too small, the model performs poorly on both the training and validation sets. As $\beta$ rises, both training and validation losses reduce greatly. Therefore, a proper $\beta$ is helpful for reaching a good local minima.

\section{Conclusion}
We present a deep VIB framework to address overfitting when training CNN classifiers on low-resource audio datasets.
We conduct extensive experiments to verify its effectiveness in removing redundant information. Although we implement VIB method on CNNs in this paper, it is also suitable with other deep learning feature extractors. Therefore, our approach has good potential in audio classification tasks.

\section{Acknowledgement}
This work is supported by National Key Research and Development Program of China under grant No.2018YFB0204403, No.2017YFB1401202 and No.2018YFB1003500. Corresponding author is Jianzong Wang from Ping An Technology (Shenzhen) Co., Ltd.

\newpage
\bibliographystyle{IEEEtran}

\bibliography{mybib}


\end{document}